\documentclass[12pt,preprint]{aastex}

\def\degree{\ifmmode {^\circ}\else {$^\circ$}\fi}
\def\rstar{\ifmmode {\, R_{\star}}\else $R_{\star}$\fi}
\def\msol{\ifmmode {\, M_{\odot}}\else $M_{\odot}$\fi}
\def\rsol{\ifmmode {\, R_{\odot}}\else $R_{\odot}$\fi}
\def\lsol{\ifmmode {\, L_{\odot}}\else $L_{\odot}$\fi}
\def\msolyr{\ifmmode {\,M_{\odot}\,{\rm yr}^{-1}}\else $M_{\odot}\,{\rm yr}^{-
1}$\fi}
\def\mdot{\ifmmode {\,\dot{M}}\else $\dot{M}$\fi}

\def\mdotyr{\ifmmode {\,\dot{M}\,yr^{-1}}\else $\dot{M}\,yr^{-1}$\fi}

\begin{document}

\title{Dust Grain-Size Distributions From MRN to MEM}
\author{Geoffrey C. Clayton$^1$, Michael J. Wolff$^2$, Ulysses J. Sofia$^3$,
K. D. Gordon$^4$ and K. A. Misselt$^4$}

\altaffiltext{1}{Department of Physics \& Astronomy, Louisiana State 
University,
   Baton Rouge, LA 70803; Email: gclayton@fenway.phys.lsu.edu}

\altaffiltext{2}{Space Science Institute, 3100 Marine Street, Ste A353,
Boulder, CO 80303-1058; Email: wolff@colorado.edu}

\altaffiltext{3}{Department of Astronomy, Whitman College,
Walla Walla, WA 99362;
E-mail: sofiauj@whitman.edu }

\altaffiltext{4}{ Steward Observatory, University of Arizona, Tucson, AZ 
85721;
E-mail: kgordon@as.arizona.edu, misselt@as.arizona.edu}

\begin{abstract}
Employing the Maximum Entropy Method algorithm, we fit
interstellar extinction measurements which span the wavelength range
0.125-3 \micron.  
We present a uniform set of MEM model fits, all using the same grain
materials, optical constants and abundance constraints.  
In addition, we are taking advantage of improved UV and IR data and better 
estimates of the gas-to-dust ratio.
The model fits cover the entire range of extinction properties that have 
been seen 
in the Galaxy and the Magellanic Clouds.
The grain models employed for this presentation are the simplistic
homogeneous spheres models (i.e., Mathis, Rumpl, \& Nordsieck 1977)
with two (graphite, silicate) or three (graphite, silicate, amorphous
carbon) components.  Though such usage is only a first step, the
results do provide interesting insight into the use of grain size as
a diagnostic of dust environment. We find that the SMC Bar extinction
curve cannot be fit using carbon grains alone. This is a challenge to the recent
observational result indicating little silicon depletion in the SMC.

\end{abstract}


\keywords{Extragalactic, extinction, dust, maximum entropy method,
Magellanic Clouds}

\section{Introduction}

It was suspected as long as 150 years ago that something was blocking the light
of stars on its way to Earth, but it wasn't until the early twentieth century
that the work of Barnard and Trumpler confirmed the existence of obscuring
clouds of interstellar dust (Whittet 1992 and references therein).  Trumpler
found that the wavelength dependence of
interstellar extinction is proportional to $\lambda^{-1}$~implying submicron
size dust grains. Despite these early advances, we, in the early 
twenty-first century, are still struggling to understand the nature of 
cosmic dust. The chemical composition of the dust grains has been the most 
difficult problem to address since the available observational data do not
provide strong constraints. Many solids have been suggested including ices,
silicates, various carbon compounds, metals and complex organic molecules. 
Graphite has been a popular grain constituent since it was
associated with the 2175 \AA\ extinction bump (Stecher \& Donn 1965). Two
silicate features in the IR are widely observed providing one of the few
firm identifications of a grain material (Woolf, \& Ney 1969; 
Gillett, \& Forrest 1973). 

The best diagnostic for determing the sizes of 
interstellar dust grains is the wavelength dependence
of extinction (Mathis 2000). A wealth of observational data now exists which
characterizes extinction properties from the UV to the IR along sightlines
in various dust grain environments. This combined with better constraints on
the abundances of elements available for condensation into grains in the
interstellar medium makes a study of
the distribution of sizes of the dust grains into perhaps a slightly 
more tractable problem than it previously was.
When data on the extinction were only available at
visible wavelengths, the proposed size distribution was flat 
up to some cutoff size (Oort \& van de Hulst 1946).
The addition of data in the UV and IR wavelength regions in the last 40 
years has revolutionized the study of the size distribution of dust grains,
putting much more stringent constraints on the upper and lower size cutoffs 
and on the slope of the size dependence.
The modern era of grain modeling was ushered in by 
Mathis, Rumpl, \& Nordsieck (1977 [MRN]).
Their grain model 
consisted of power-law size distributions of separate populations
of bare spherical silicate and graphite grains.  
It is a testament to the groundbreaking nature of this work that today,
twenty-five years later, it has not been fully superseded by later studies.

MRN-type analyses
must make assumptions regarding a particular model of the
size-dependence of the dust grains.  One is now able
to circumvent such issues through a solution of the inverse problem
using the Maximum Entropy Method (MEM) (e.g., Kim, Martin, \& Hendry 1994).  
The versatility of this technique allows one to easily
include a range of grain models in the analysis: from the simple
MRN-like approach to more realistic composite grains which
can include the effects of porosity, solid inclusions (silicate,
graphite, superparamagnetic etc.), and alignment.
In this work, we present some preliminary efforts to 
extract MEM size distributions
of dust grains in the Local Group using ``basis sets'' of
grain cross-sections for several composite grain models, each of
which satisfies cosmic abundance constraints.

\section{History of Grain Size Distributions}
Most studies of grain size over the last half century have assumed a size 
distribution
which is either a
power law or a power law with exponential decay (PED).  
In addition, most calculations have used Mie Theory assuming spherical grains.  
These power laws
have been utilized for two reasons. First, they provide a good fit to the observed 
extinction 
wavelength dependence with as few as two populations of grains. Second, formation 
and destruction 
processes such as shattering and coagulation which
modify grain sizes along many sightlines may naturally produce a power law size 
distribution 
(Biermann \& Harwit 1980; Hayakawa \& Hayakawa 1988; O'Donnell \& Mathis 1997).
This type of study began with
Oort \& van de Hulst (1946) who fit extinction at visible wavelengths assuming ice 
grains and a
size distribution resembling a PED. Greenberg (1968) greatly expanded this work 
using spheres,
cylinders and spheroids, combined with many different materials and grain orientations limited 
only by lack of 
computing power available at  the time. Greenberg utilized the Oort \& van de 
Hulst size
distribution almost 
exclusively. 

This brings us to MRN who found that a very good fit could be made to the UV and 
visible 
extinction using
separate populations of bare silicate and graphite grains with a power law 
distribution of sizes where 
n(a) $\propto$
$a^{-3.5}$. The graphite grains range in size from 0.005 to 1 \micron~and the 
silicate grains from 
0.025 to 0.25 \micron.  
Mathis \& Wallenhorst (1981) expanded on MRN by showing that they could fit 
various high $R_V$\footnotemark
\footnotetext{$R_V$ is the ratio of total-to-selective extinction and is a crude 
measure of the average size 
of the dust grains along a particular line of sight.}
 sightlines
with ``peculiar" extinctions such as $\sigma$ Sco, $\rho$ Oph
and $\theta$ Ori by
raising the lower size cutoff to 0.04 \micron~and sometimes also raising the upper 
size cutoff.  This suggested that
small silicates
are missing along these lines of sight.
Greenberg \& Chlewicki (1983) claimed that grains responsible for bump and 
FUV extinction 
are two 
independent populations and that MRN violates this condition.
However, Cardelli, Clayton, \& Mathis (1989 [CCM]) showed that the Greenberg \& 
Chlewicki result is 
spurious due to having an
$R_V$ dependence in
one ratio and not in the other.

The lack of accurate optical constants for silicates and graphite have cast doubts 
on the 
significance of the good fits to extinction data with MRN-type models. 
In the absence of laboratory data on graphite, MRN adjusted the optical constants 
to 
make the 
bump lie at 2175 A. The situation improved with the work of Draine \& Lee (1984) 
who
synthesized dielectric functions for graphite and ``astronomical" silicate on the 
assumptions that MRN is valid. The polycyclic
aromatic hydrocarbon (PAH)
absorption cross-section was established based on the interstellar 2175 \AA\ bump 
profile (Li \& Draine 2001).
As data from the mid- and far-IR became available, it became evident that 
significant numbers of small grains
and large molecules such as PAHs were needed to account for emission features and 
excess continuum
emission from single-photon heating (Draine \& Anderson 1985; Desert, Boulanger, 
\& Puget 1990).
The IRAS data imply a need to push the lower size cutoff of MRN down to 
$\sim$0.0003 \micron.

CCM suggested that all populations of 
grains are subject to the same processes in a systematic 
way across the whole size distribution.
They found that there is an average Milky Way extinction relation, 
A($\lambda$)/A(V), 
over the wavelength range 0.125 $\mu$m to 3.5 $\mu$m, which is
applicable to a wide range of interstellar dust environments,
including lines of sight through diffuse dust, dark cloud dust, and 
star formation (CCM; Cardelli \& Clayton 1991; Mathis 
\& Cardelli 1992).
The extinction relation is a function of one parameter, chosen to be R$_V$.
The existence of this relation, valid over a large wavelength
interval, suggests that the environmental processes which modify the
grains are efficient and affect all grains.
Mathis \& Whiffen (1989) suggested that many grains are composite, perhaps 
like 
interplanetary grains
with many individual grains of different materials including silicates and 
amorphous carbon 
stuck together in grains 
with significant amounts of vacuum.  In their model, 
small bare graphite grains are still needed 
to explain the bump
and PAHs for the IR.

MRN-type fits and the Draine \& Lee optical constants have also been applied 
to 
sightlines in the
the Magellanic 
Clouds. 
Pei (1992) constructed one characteristic curve for each of 
the Galaxy, LMC and SMC even though large variations are seen from sightline to 
sightline in the SMC
and LMC let alone the Milky Way. He used data from Koornneef (1982, 1983), Rieke 
\& Lebofsky (1985),
Morgan \& Nandy (1982), Nandy, Morgan, \& Houziaux (1984) and Bouchet et al. 
(1985). This study did 
not use the best
extinction curve data available even at that time. The wavelength dependence of UV 
extinction in the
LMC and SMC are much better characterized now (Gordon \& 
Clayton 1998; 
Misselt, Clayton, \& Gordon 1999; Gordon et al. 2002). 
Pei (1992) uses the same exponent, and upper and lower cutoffs as MRN then varies 
the relative
abundances of carbon and silicon in the three galaxies to get the best fit. 
He finds minor differences between the best fits and the measured 
curves, and suggests that changing the exponent and upper and lower cutoffs
wouldn't improve on the fits presented. Pei finds it,``remarkable that all three 
parameters are 
roughly optimized for
all three galaxies."
The standard silicate/graphite model appears to account well for all of Pei's 
average curves
with only the adjustment of the graphite to silicate ratio.

Aannestad (1995) expanded on MRN and extended fitting the extinction curve out to 
925 \AA\
using Voyager data.  He used both Mie calculations for spheres plus 
the discrete dipole approximation (DDA) for graphite disks.
Large and small silicates and carbon grains, and  PAHs were included along with
organic refractory mantles (Jenniskens 1993). Aannestad suggests that the very 
small grains are made up of
graphite, amorphous carbon, and diamond.
As with previous studies, the higher $R_V$ extinction curves are fit
by eliminating small grains. 
He also finds that 2:1 oblate graphite disks are better than spherical graphite 
for fitting the bump.
Aannestad suggests that radiation pressure could be responsible 
for changing grain
size distributions in addition to shattering and coagulation.

Aannestad's study along with two that followed it (Mathis 1996; Li \& Greenberg 
1997) represent attempts
to update MRN by addressing the complexities of fitting the UV to the far-IR 
extinction 
while staying
within the constraint of the available abundances of various elements.  The latter 
two studies
use a PED for defining the size distribution. 
Mathis (1996) uses Mie theory but doesn't attempt to include very small grains.
Li \& Greenberg (1997) employ finite cylinders for modeling the core-mantle 
grains. 
They state that their model uses less carbon than other models and so isn't so 
constrained by the carbon abundance.

Dwek et al.~(1997) fit dust models to IR emission from 3.5 to 1000 \micron~using
IRAS and COBE data.
The IRAS data first showed cirrus emission with an excess at 12 and 25 \micron~
caused by fluctuations in small grains down to 3 \AA\ in size.
These may be the same as the PAHs that cause UIR bands. 
The Dwek et al. model uses bare silicate and graphite grains and PAHs.
They make a fit to 14 parameters, 8 fixed and 6 varying,
using Mie theory with the Draine \& Lee constants.
Once the IR emission was fit they calculated the predicted UV extinction for their 
derived
size distribution. 

In addition to the many MRN-like studies outlined above, two other methods 
have 
been suggested for 
deriving the dust grain size distribution by solving the inverse problem. 
Zubko, Krelowski, \& Wegner  (1996a, 1998) suggest regularization where once a 
grain model is assumed then a 
unique 
size distribution 
is derived. It is not restricted by any a priori relation such as a power law.
However, like most of the other studies above, Zubko et al. use bare spherical 
grains (astronomical 
silicate, graphite, AC, SiC and water ice) and Mie theory to calculate the 
extinction efficiencies.
Their best fit models are silicate/graphite/ice mixtures 
although, as they note, the 3.1 
\micron\ ice 
feature is not seen in diffuse dust.
Zubko et al. say the results are essentially different from the MRN power law. His 
method results in size 
distributions 
with more structure than a simple power law. They are qualitatively similar to the 
results of the MEM method 
described below. 
The regularization method has also been applied to the SMC extinction (Zubko 
1999). 

The other derivation of the size distribution is through the Maximum Entropy 
Method (MEM). Kim et al. (1994)
use the MEM method with MRN-type bare spherical silicate, graphite and amorphous 
carbon particles for two 
cases 
with $R_V$ = 3.1 and 5.3.
The main constraint applied is the 
wavelength
dependence of interstellar extinction.
A second constraint is the abundance of various elements. This is combined with
an assumed gas-to-dust ratio, $N_H/A_V$.
Their results are similar to MRN but the resulting size distribution has more 
structure.
The upper size cutoff in MRN was a subjective choice. MEM results show a smooth 
decrease starting at 0.2 \micron.
The amounts of carbon and silicon used are similar to MRN. 
Solar abundances as defined by Grevesse \& Anders (1989) were used to constrain 
their models. 
Kim \& Martin (1996) expanded the MEM work to a wider range of $R_V$ values.  
They fit $R_V$ = 2.7 to 5.3, while trying to minimize amount of carbon 
used.
They limited the mass to that used by $R_V$=3.1.
They also found that fitting the extinction curve for $\rho$ Oph needs more large 
particles in a 
bimodal distribution.

Recently, Weingartner \& Draine (2001) expanded on MRN by establishing a multifunctional 
form of the grain 
sizes including two log-normal size distributions, PEDs and curvature terms.  It 
involves
6 adjustable parameters for carbon grains and five for silicates. The fits are 
also constrained
by $R_V$ and the available carbon abundance. They fit extinction curves by varying 
the powers, the 
curvature parameters, the transition sizes, the upper cutoff parameters and the 
total volume per H atom
in both the carbon and silicate grain distributions, respectively. They include 
sufficient
numbers of small grains to account for observations of IR emission. 
They find carbon and silicate grain size distributions that can account for both 
the UV to IR extinction and 
the IR emission. 

\section{Elemental Abundances in the ISM}

An important constraint on the composition of interstellar grains is
the cosmic abundance of various elements.  A proposed grain
constituent cannot use more of any element than is available in the
interstellar medium.  
So a set of reference abundances, that represent the typical ISM in a given
galaxy, is needed. This 
set
represents the total abundance of elements in the gas and dust. If these reference 
values are known, then a 
measurement of the gas phase abundances will reveal how much is missing, i.e., how 
much has condensed 
to form dust grains.
In the last few years, new data have put much more 
stringent
constraints on the reference abundances in the interstellar medium of the Galaxy 
(e.g.,
Spitzer \& Fitzpatrick 1993; Cardelli et al. 1993, 1994; Sofia, Cardelli, \& 
Savage 1994; Snow \& Witt 1996).  In particular, there was a move away from the 
use of 
solar abundances.  This was motivated by the fact that solar abundances 
implied more
oxygen than could be accounted for by the gas and solid phases of the ISM. 
As a result, it was suggested that all abundances may be about 60\% of Solar. 
Other abundance sources such as B stars have been used instead.
One impact 
of this has been the so-called ``carbon crisis'', in which
there may be less carbon available than is needed for various models of the 
interstellar extinction 
including MRN.  In fact,
Mathis (1996) shows this paucity of carbon would likely play a large
role in constructing and validating grain models.  
Recently, however, it has become apparent that that the ``crisis'' may be due to 
problems with the 
solar abundances. The measurements of the abundance of oxygen in the Sun have 
been dropping 
and now stand at about 
two thirds of its former value (Sofia \& Meyer 2001). 
There is now no good argument for adopting a reference that is two-thirds the 
solar 
abundance (Sofia \& Meyer 2001).
Therefore, Solar abundances have been adopted as the standard for the the Galactic 
sightlines in
this study. See Table 1.

In the Magellanic Clouds, the relative abundances between the elements are similar 
to those found 
in the local ISM (Russell \& Dopita 1992; Welty et al. 1997; Venn 1999).  However, 
the 
metallicities in the LMC and SMC are 0.3 and 0.6 -- 0.7 dex lower, 
respectively,
than 
in the local ISM in the Galaxy (Welty et al. 2001). Metallicity and the gas-to-dust
ratio are correlated, so the amount of dust per hydrogen column is lower in the 
Magellanic Clouds.  

The observed gas-phase abundances of various elements in the ISM should be just 
the reference abundance 
minus any depletion into dust grains.  In the 
Galaxy, several depletion patterns have been associated with different 
environments (Savage \& Sembach 1996; Welty et al. 2001).  
For example, the depletions range from -0.8 [Mg, Si/Zn] and  -1.8 [Fe, Ni/Zn] to -
0.2 [Mg, Si/Zn] 
and -0.5 [Fe, Ni/Zn] for cold dense clouds and the Galactic halo, respectively, 
with the warm diffuse 
ISM depletions lying in between.  The relative gas-phase abundances in the 
Magellanic clouds resemble 
those in the Galactic halo implying that the depletion patterns in the ISM of 
those galaxies may be similar to those in the Galaxy despite the known differences 
in 
gas-to-dust ratio and metallicity (Welty et al. 2001).  However, for at least 
three SMC 
sightlines, the depletion pattern shows one important difference.  Silicon appears 
to be 
almost undepleted ($<$ 0.2 dex).  Almost every grain model for the ISM from MRN to 
the 
present time includes silicates as an in important constituent.  In particular, 
grain 
models of the SMC have emphasized the importance of silicates since the role of 
carbon 
seems to be less important as measured by the strength of the 2175 \AA\ feature 
(Pei 1992; 
Zubko 1999; Weingartner \& Draine 2001).  So, if silicon is indeed relatively 
undepleted in the SMC, 
the grain models will have to be reassessed (Welty et al. 2001).

\section{Extinction Curves} 
Low dispersion short and long wavelength IUE spectra, combined with BVJHK 
photometry
were used to construct the extinction curves in this study.  The spectra 
were downloaded from the MAST IUE archive.  The archive spectra were reduced using
NEWSIPS and then were recalibrated using the method developed by Massa \&
Fitzpatrick (2000).  The short and long wavelength spectra for each star were
co-added, binned to the instrumental resolution ($\sim$5 \AA) and merged at the
maximum wavelength of the short wavelength spectrum.
Extinction curves were constructed using the standard pair method (e.g., Massa,
Savage \& Fitzpatrick 1983).  Uncertainties in the extinction curves contain terms
that depend both on the broadband photometric uncertainties as well as
those in the IUE fluxes, which are
calculated directly in NEWSIPS.  Our error analysis is described in detail in
Gordon \& Clayton (1998).  The sample includes early-type supergiants which may be
used with the same accuracy as main sequence stars in calculating extinction
(Cardelli, Sembach, \& Mathis 1992).  Table 2 lists the sightlines being analyzed 
for this study.

The extinction curves have been fitted using the Fitzpatrick \& Massa (1990,
hereafter FM) parameterization.  FM have developed an analytical representation
of the shape of the extinction curves using a small number of parameters.  This 
was
done using linear combinations of a Drude bump profile, $D(x;\gamma,x_o)$, a 
linear
background and a far-UV curvature function, $F(x)$, where $x = \lambda^{-1}$.
There are 6 parameters determined in the fit:  The strength, central wavelength,
and width of the bump, $c_3$, $x_o$, and $\gamma$, the slope and intercept of the
linear background, $c_1$ and $c_2$, and the strength of the far-UV curvature,
$c_4$.  The FM best fit parameters to the sightlines  
are given 
in Table 3.

Extinction data have been obtained from the literature for the LMC, SMC and some 
Galactic
sightlines (Gordon \& Clayton 1998; Misselt et al. 1999; Clayton, Gordon, \& Wolff 
2000). 
Various ``average''
curves have been calculated which generally represent extinction characteristics 
of groups 
of sightlines.
In the SMC, only four suitable early-type stars have been observed that are
significantly reddened and also had well matched comparison stars (Gordon
\& Clayton 1998). Three stars lie
in the SMC Bar, and the line of sight for each of them passes through regions
of recent star formation.  The fourth star belongs to the SMC Wing and its
line-of-sight passes though a much more quiescent region.
In the LMC, the sample contains 12 stars in the 30 Dor region
and 7 stars outside 30 Dor (Misselt et al. 1999).  While the extinction properties 
of the 30
Dor and non-30 Dor samples as defined are both inhomogeneous,
a group of stars with similar reddenings and bump strengths
were found lying in or near the region occupied by the supergiant
shell (LMC 2).  The average extinction
curves inside and outside LMC 2 show a very significant difference in 2175 \AA\ 
bump strength, but the 
far-UV extinctions are very similar.

An ``average'' extinction curve for the Galaxy is included in Tables 2 and 3. It is representative 
of a typical 
diffuse interstellar sightline. It has been constructed by making an FM fit to a 
CCM curve with 
$R_V$=3.1 (Misselt et al. 1999). Another average curve is included which 
represents dust 
characteristics along extremely long path length, low-density sightlines in the 
Galaxy from the sample of Sembach \& Danks (1994, SD region)
(Clayton et al. 2000).
In addition, several individual sightlines are included in this study which are representative 
of the range of extinction 
characteristics seen in the Galaxy. They are listed in Table 2.  
These extinction curves have been constructed as described above 
(Valencic et al. 2003). Extinction curves have been presented for these stars 
previously: HD 29647, 
HD 62542 (Cardelli
\& Savage 1988); HD 37022, HD 147889, HD 204827 (Fitzpatrick \& Massa 1990); HD 
147165 
(Clayton \& Hanson 1993); HD 210121 (Larson,
Whittet, \& Hough 1996); 
and the SD region (Clayton et al. 2000).
These sightlines
comprise values R$_V$ from 2.1 to 5.5. 
The sample includes the ``standard" 
CCM curve ($R_V$ = 3.1) as well as those with steep (HD 210121, SD Region, 
HD 204827) and flat (HD 37022, HD 147165, HD 147889) UV rises, normal
and weak (HD 29647, HD 62542) bumps. These sightlines represent all the 
physical 
environments accessible to UV observations including dense clouds, 
star-forming 
regions and the diffuse ISM.  In addition, we include average sightlines for 
the 
two extinction-curve types seen in each of the SMC and LMC.

\section{Maximum Entropy Method}

We employ a (slightly) modified version of the MEM extinction
fitting algorithm developed by Kim et al. (1994, and
references within).
Instead of using the number of grains as a constraint, the algorithm employs
the mass distribution:  m(a) da = mass of dust grains per H atom in the size
interval a to a+da.
Thus, the traditional MRN-type model becomes $m(a) \propto a^{-0.5}$.
We use a PED as the template
function for each component.
The data are examined at 34 wavelengths and the grain cross
sections are computed over the range 0.0025-2.7 \micron\ with 50
logarithmically-spaced bins.
The shape of the mass distribution is strongly constrained only for data over
the region ~0.04-1 \micron.  Below 0.04 \micron, the Rayleigh scattering
behavior constrains only total mass;
above 1 \micron, the ``gray'' nature of the dust opacity also
forces the MEM algorithm to simply adjust the total mass, using the
shape of the template function to specify the size-dependence of the
distribution.

The total mass of dust is constrained using both the gas-to-dust ratio
and ``cosmic" abundances (i.e., we try not to use more carbon or silicon 
than is
available); see Table 1.
As a first step, we consider only  three-component
models of homogeneous, spherical grains: modified ``astronomical silicate"
(Weingartner \& Draine 2001), amorphous carbon (Zubko et al. 1996b), and
graphite (Laor \& Draine 1993).

We modeled all the individual and average sightlines for the Galaxy and the 
Magellanic Clouds listed in Table 2. The results are presented in Figures 1-3. For 
each sightline, two plots have been produced. The first shows the amount of 
extinction provided by each of the three materials, the total extinction of the 
model compared to the measured extinction curve. The fraction of the adopted 
abundance available for each material (Table 1) which is needed for the best model 
fit, is also listed. The second plot shows the resultant mass distribution for 
different sizes of grains of each material relative to the mass of hydrogen.

Figures 1 and 2 present the three-component models for a selection of Galactic 
sightlines including
the ``standard" curve ($R_V$ = 3.1). 
Figure 3 presents  three-component models for the SMC and
LMC curves described above.

\section{Discussion}

The MEM fits to all of the sightlines considered here show the same general behavior with 
the dust grain 
mass distributed 
across a 
wide range of sizes up to some upper size cutoff.
As described above, the MEM model is not well constrained for very large 
and very 
small grains. 
One would naively think that as $R_V$ gets larger and the UV 
extinction gets flatter that the relative amount of mass in small grains 
would go 
down and large grains would go up; for steeper curves, the opposite would happen.
These effects are seen in 
Figures 1-3 but no simple correlation with $R_V$ or far-UV steepness is seen.  
Similarly, sightlines with weak bumps use relatively less mass in the 
likely bump grains. One trend is that sightlines with higher than 
average gas-to-dust ratios lie below the average Galactic curve at all grain sizes.  This 
is just due to the fact that the the masses in the figures are normalized to 
the mass of hydrogen.  

The fraction of the available silicon and carbon used in the 
MEM fits to the various sightlines covers  a very wide range.  
Three general factors determine the fraction of silicon and carbon
that any individual sightline will use. 
First, the higher the gas-to-dust ratio is, the more metals are available 
in the gas phase.
Second, the higher the abundances of metals are, the more material is available.
Finally, high values of $R_V$ imply a greater than average mass fraction in larger
grains. The extinction due to large grains is not as efficient per unit mass as
smaller grains since the mass goes as $a^3$ while the surface area 
goes as $a^2$.  
HD 37022 which has normal Galactic abundances and gas-to-dust ratio but the highest
value of $R_V$ in our sample, requires more than 100\% of the available silicon 
and carbon. 
Similarly, the other high $R_V$ sightlines, HD 147889, HD 147165 and HD 29647 all 
require large fractions of the available metals. The SMC Wing which has a 
low gas-to-dust
ratio and low abundances also uses more than 100\% of the available silicon 
and 
carbon.
On the other hand, the SD region with a high gas-to-dust
ratio, normal abundances and a low $R_V$ uses less than half of the silicon 
and 
carbon. The LMC2 and the SMC Bar regions both have low abundances and 
high gas-to-dust
ratios. Both use 50\% or less of the available silicon and 60-80\% of the carbon. 
The SMC Bar uses a larger fraction of both elements because of its higher 
normalized 
extinction, requiring more dust grains.
The relationship between elemental fraction used and the gas-to-dust ratio is illustrated in 
Figure 4.
 
All extinction models rely heavily 
on silicates, yet recent observations suggest that 
silicon is relatively 
undepleted in the SMC Bar
(Welty et al. 2001).
It should be noted that the sightlines investigated by Welty et al. do not 
have extinction curves due to their small reddenings.
However, MEM model fits to the average SMC Bar extinction 
indicate that the curve cannot be fit with carbon grains
alone.
Since there is almost no bump in the SMC Bar extinction that may also imply the 
presence of fewer graphite grains.  
Carbon grains are responsible for most of the visible extinction and silicon grains 
for most of the UV extinction. So, in general, both species of grains are needed
along any sightline to get a good fit to the extinction curve. 
At least, the MEM fit for the SMC Bar requires a smaller fraction of 
silicon and carbon 
than almost any other sightline modeled.

A quick look at the figures shows that the HD 210121 and SMC Bar sightlines are 
the most unusual in their grain 
size distributions. 
Both differ strongly from the CCM curve. 
These two sightlines also have the steepest far-UV extinction. The model fits show a 
deficit of large silicate and amorphous carbon grains for both sightlines.   
The HD 210121 sightline which, unlike the SMC Bar, has a significant bump, shows 
a slightly larger size cutoff for graphite grains. The next steepest sightlines, in 
the LMC and in the SD region in the Galaxy do not show much difference from the 
average curve once the gas-to-dust ratio differences are taken into account. 
The importance of small grains in the steep UV
rise for several sightlines (including HD 204827) is also quite evident.
Larson et 
al. (2000) also did MEM modeling for HD 210121 with similar results.
The largest $R_V$ sightline, toward HD 37022, also shows large deviations from 
the average Galactic curve. The MEM fit to the HD 37022 sightline shows a 
large 
deficit of small graphite and silicate grains. 
Kim et al. (1994) found that extinction with $R_V$=5.3 requires a similar dust
mass to extinction with 
$R_V$=3.1. There is a 
great reduction in 
the number
of grains with sizes less than 0.1 \micron\ and some increase in larger particles 
for the model with the 
larger value of $R_V$.
Among the fits to other large $R_V$ sightlines, HD 147165 shows 
trends similar to the HD 37022 fit  while the fit to HD 147889 is not 
significantly different from
the average Galactic fit.
The mass distributions, found in this study, show significant departures
from power-law (and even PED) behavior.  This is an important distinction
to make in light of the work by Pei (1992), which extends MRN to the same
environments by changing only the end points of the grain size range.

As an interesting exercise, we constructed a composite model along
the lines of Mathis (1998).  However, given our interest in increasing
the cross section per unit volume while also minimizing the amounts
of Si and C used, we chose to explore the use of pyroxene and oxides.
In other words, by freeing up some of the [Fe,Mg] normally used in
olivine-like astronomical silicate, one is able to form oxides without
consuming any additional cosmic Fe or Mg (as compared to olivine).
Briefly, our composite grain component consists of (by volume)
28.5\% pyroxene ($[Fe_{0.6}Mg_{0.4}]SiO_3$), 16.5\% amorphous
carbon, 5\% oxide ($[Fe_{0.4}Mg_{0.6}]O$), and 50\% vacuum.  The
elemental budget for this component is [C/H] = 1.05e-4 and
[Si/H] = 2.9e-5.  Defining a ``molecule" of this material to
have 1 Si atom, one derives a molecular weight of 213 and density of
1.45 g/cm$^3$.
The dust mass densities used in this study are $\rho_{sil}$= 3.3 g~$cm^{-3}$, 
$\rho_{gra}$= 2.3 g~$cm^{-3}$ and $\rho_{amc}$= 1.8 g~$cm^{-3}$.
We included small silicates and small
carbon (graphite) grains in our composite model, in addition
to the composite grain component itself.

Following the lead of Mathis (1998), we computed the optical properties using a
geometric mean of two effective medium mixing rules:
Bruggeman and an extension by Ossenkopf (1991).  The dielectric
functions for the individual components were taken from the following
sources:  amorphous carbon - Be form of Zubko et al. (1996b); 
oxide - Henning et al. (1995), and pyroxene - Dorschner et al. (1995).
Unfortunately, the latter two are limited to wavelengths above 0.22 \micron\
(we discarded the 0.2 \micron~point).
For the purpose of this exercise, we assumed that the UV behvior
of the materials will be similar to that of astronomical silicates
(i.e., the Fe component is dominant in this regime).  So,
we extended the dielectric function to the 0.08-0.2 \micron\ region
by scaling the imaginary component ($k$) of astronomical silicates
to match smoothly with that of each material at 0.22 \micron.  We
then computed the real component ($n$) using a subtractive Kramers-Kronig
(SKK) algorithm, implementation kindly supplied by Kelly Snook
(Snook 1999; see also Warren 1984).  
The fixed-point value of $n$ needed for the SKK calculations is adopted
from the visible measurements of each material at a wavelength of 0.5 \micron.
The model fits are shown in Figure 5. This result is intriguing as it indicates that
the composite grains may require a smaller fraction of the available silicon and
carbon.

Weingartner \& Draine (2001) note that Kim et al. (1994) make more efficient use of the 
grain volume by using MEM to 
construct more complicated size distributions.  They find this ``fine-tuning 
unappealing.'' 
However, their own models have trouble 
staying within the standard abundance/depletion 
limits.
Weingartner \& Draine suggest that it is
more 
likely that the abundance constraints are too stringent and the real Galactic 
abundances are higher than 
solar.
As can be seen in Figures 1 and 3 of this paper, the MEM model fits sometimes
require 100\%  or more of the available silicon and carbon assuming the ``new" 
old solar abundances.
Weingartner \& Draine fit the Galactic average extinction curve as well as
the average LMC, LMC2 and SMC Bar sightlines. A comparison of their results with 
ours shows significant differences, in particular, in the upper size cutoffs of 
silicate and carbon grains. 
For example, Weingartner \& Draine also fit the HD 210121 sightline. Their two-component fit 
has upper size cutoffs of $\sim$0.3 \micron~for the silicates and $\sim$1 
\micron~for carbonaceous grains.  In our three-component fit, we find the opposite, 
$\sim$0.3 \micron~for amorphous carbon and $\sim$1 \micron~for silicate grains.  
We also include graphite with an upper cutoff of $\sim$0.3 \micron.
Dwek et al. 
(1997) used IR emission at mid-IR wavelengths to constrain the size distribution 
and then predicted the UV extinction.
They find a size distribution for silicates not too different from MRN and 
include 
two power laws for graphite.
Their predicted UV extinction curve is steeper in the UV than the $R_V$=3.1 
extinction curve. The cirrus, modeled by Dwek et al. (1997), is located at
high latitude so may have extinction  more like that seen toward HD 210121.
The differences among these studies reinforce the fact that while the model results are useful, 
they are not unique.
Li \& Greenberg (1998) fit the extinction curve of HD 210121 using both
an MRN power-law with silicates and graphite, and a core-mantle model.

Most previous papers that calculate the size distributions assume just one 
``average" extinction curve to fit (i.e., Mathis 1996).   
In this paper, we present a uniform set of MEM model fits, all using the same grain
materials, optical constants and abundance constraints.  
In addition, we are taking advantage of improved UV and IR data and better 
estimates of the gas-to-dust ratio.
The model fits cover the entire range of extinction properties that have been 
seen 
in the Galaxy and the Magellanic Clouds.

Clearly, more work needs to be done:
The derived mass distributions will need to be tested in other
observational regimes, such as thermal IR and linear polarization.
More sophisticated cross sections need to be included, both in
terms of grain topology and composition.  The next step will include
composite grains, as well as both aligned and randomly-oriented spheroids.
In addition, more extensive Galactic studies need to be undertaken.
Specifically, attempts to tie sightline characteristics such as depletion
patterns, molecular abundances, etc. to specific extinction features (i.e.,
steep UV rise, weak bump).

\acknowledgements
We would like to thank Kelly Snook for her help with this project. This work was
suported by NASA ATP grant NAG5-9203.

\clearpage
%
%
\begin{deluxetable}{lccccl} 
\tablecolumns{6} 
\tablewidth{0pc} 
\tablecaption{Adopted Abundances. }
\tablehead{ 
\colhead{}    &  \multicolumn{2}{c}{``Cosmic''} & 
\multicolumn{2}{c}{In Dust} \\ 
\colhead{} & \colhead{Si/H}   & \colhead{C/H}    & \colhead{Si/H} & 
\colhead{C/H} & \colhead{References}}
\startdata 
Galactic        &4.0e-5&3.2e-4&3.8e-5&1.8e-4&1 \\
Galactic (Halo) &4.0e-5&3.2e-4&2.0e-5&1.8e-4&1 \\
SMC             &1.1e-5&5.4e-5&1.1e-5&5.4e-5&2,3 \\
LMC             &6.5e-5&1.1e-4&6.5e-5&1.1e-4&2,4 \\
\enddata 
\tablecomments{ 1. Sofia \& Meyer 2001, 2. Russell \& Dopita 1992,
3. Welty et al. 1997, 4. Welty et al. 1999}
\end{deluxetable} 

%
%
\begin{deluxetable}{lcccl} 
\tablecolumns{5} 
\tablewidth{0pc} 
\tablecaption{Observational Characteristics of the Modeled Sightlines. }
\tablehead{ 
\colhead{Sightline} & \colhead{$R_V$} & \colhead{$E_{B-V}$} &
\colhead{$N_H/E_{B-V}$} & \colhead{References$^a$}}
\startdata 
SMC Bar  &2.7&0.18&3.6e+22\tablenotemark{b}&1,2,18 \\
SMC Wing &2.1&0.24&1.5e+21\tablenotemark{b}&1,3,18 \\
LMC      &3.4&0.25&1.1e+22&4,5,18 \\
LMC 2    &2.8&0.19&1.9e+22&4,5,18\\
Average Galactic&3.1& \nodata &5.8e+21&6,7 \\
HD 29647&3.6&1.00&5.8e+21&6,12\\
HD 37022&5.5&0.34&5.8e+21&6,7\\
HD 62542 &3.2&0.33&5.8e+21&6,11,13 \\
HD 147165&3.8&0.40&5.9e+21\tablenotemark{b}&7,14,15\\
HD 147889&4.2&1.09&5.8e+21&7\\
HD 204827&2.6&1.10&5.8e+21&6,8 \\
HD 210121&2.1&0.38&5.8e+21&9,10\\
SD Region\tablenotemark{c}&2.7&0.24&1.2e+22\tablenotemark{b}&16,17\\
\enddata 

\tablenotetext{a}{ (1) Gordon \& Clayton 1998, (2) Martin et al. 1989,
(3) Fitzpatrick 1985, (4) Gunderson et al. 1998,
(5) Misselt et al. 1999, (6) Bohlin et al. 1978, (7) CCM,
(8) Clayton et al. 1995, (9) Larson et al. 1996, (10) Larson et al. 2000, (11) 
Cardelli \& 
Savage 1988, (12) Whittet et al. 2001, (13) Whittet et al. 1993, (14) Clayton \& 
Hanson 1993,
(15) Savage et al. 1977, (16) Clayton et al. 2000, (17) Sembach \& Danks 1994, 
(18) Gordon et al. 2002}
\tablenotetext{b}{$N_{HI}/E_{B-V}$}
\tablenotetext{c}{Average of 7 sightlines (Clayton et al. 2000)}

\end{deluxetable}

\begin{deluxetable}{lccccccc}
\tabletypesize{\scriptsize}
\tablewidth{0pt}
\tablecaption{FM Parameters}
\tablehead{\colhead{Sightline} & \colhead{$c_1$} & \colhead{$c_2$} & 
           \colhead{$c_3$} & \colhead{$c_4$} &
           \colhead{$x_o$} & \colhead{$\gamma$} & \colhead{Ref.}}
\startdata
SMC Bar & -4.47 $\pm$ 0.19&  2.35 $\pm$ 0.18&  0.08 $\pm$ 0.01&  -0.22 $\pm$ 
0.004&  --&--&1\nl         
SMC Wing& -1.00 $\pm$ 0.15&  1.17 $\pm$ 0.12&  3.49 $\pm$ 0.19&  0.17 $\pm$ 0.04&  
4.69 $\pm$ 0.0001& 1.17 $\pm$ 0.05&1\nl
LMC &  -1.28 $\pm$ 0.34& 1.11 $\pm$ 0.10 & 2.73 $\pm$ 0.37 & 0.64 $\pm$ 0.06& 
4.596 $\pm$ 0.017&  0.91 $\pm$ 0.05&2\nl
LMC 2  & -2.16 $\pm$ 0.36&  1.31 $\pm$ 0.08 & 1.92 $\pm$ 0.23& 0.42 $\pm$ 0.08&  
4.626 $\pm$  0.010& 1.05 $\pm$ 0.07&2\nl
Milky Way&0.12 $\pm$ 0.11&0.63 $\pm$ 0.04&3.26 $\pm$ 0.11&0.41 $\pm$ 0.02&4.596 
$\pm$ 0.002&0.96 $\pm$ 0.01&2\nl
HD 29647& -0.02 $\pm$  0.20 &  0.82 $\pm$ 0.02  &  3.06 $\pm$ 0.26 &   0.61 $\pm$ 
0.25 &   4.64 $\pm$ 0.03  &   1.45 $\pm$ 0.003&4\nl
HD 37022 & 2.06  $\pm$ 0.35 &  0.004  $\pm$ 0.003 &  2.14 $\pm$ 0.06 &   0.53  
$\pm$ 0.04 &  4.56 $\pm$ 0.002  &  1.04 $\pm$ 0.02&4\nl
HD 62542 &-1.31  $\pm$ 0.06 &  1.31 $\pm$ 0.09  &  2.64  $\pm$ 0.10 &  1.36  $\pm$ 
0.12 &  4.61  $\pm$ 0.01 &  1.37 $\pm$ 0.02&4\nl
HD 147165& 1.07  $\pm$ 0.20 &  0.31 $\pm$ 0.02 &   4.03 $\pm$ 0.12 &  0.04   $\pm$ 
0.003&  4.59   $\pm$ 0.002 & 1.09 $\pm$ 0.02&4\nl
HD 147889&1.89  $\pm$ 0.20 &  0.09 $\pm$ 0.018  &  4.77 $\pm$ 0.05 &   0.92  $\pm$ 
0.04 &  4.62  $\pm$ 0.01 & 1.08 $\pm$ 0.01 &4\nl 
HD 204827&-0.74  $\pm$ 0.08  & 1.10  $\pm$ 0.01 &  3.11   $\pm$ 0.11&  0.86  $\pm$ 
0.16 &  4.60   $\pm$ 0.01 & 1.16 $\pm$ 0.02&4\nl
HD 210121&-3.15  $\pm$ 0.09 &  1.96 $\pm$ 0.15  &  2.94  $\pm$ 0.15 &  1.05   
$\pm$ 0.11 & 4.57 $\pm$ 0.01 &   1.11 $\pm$ 0.02&4\nl
SD Region &-1.83 $\pm$ 0.14 & $1.19$ $\pm$ 0.20 & $1.52$ $\pm$ 0.27 & $0.58$ $\pm$ 
0.10 & $4.60$ $\pm$ 0.02 & $0.93$ $\pm$ 0.02 &3\nl
\enddata
\tablecomments{(1) Gordon et al. 2002, (2) Misselt et al. 1999, (3) Clayton et al. 
2000, (4) Valencic et al. 2002}
\end{deluxetable}

\clearpage
\begin{figure*}
\epsscale{1.0}
\plottwo{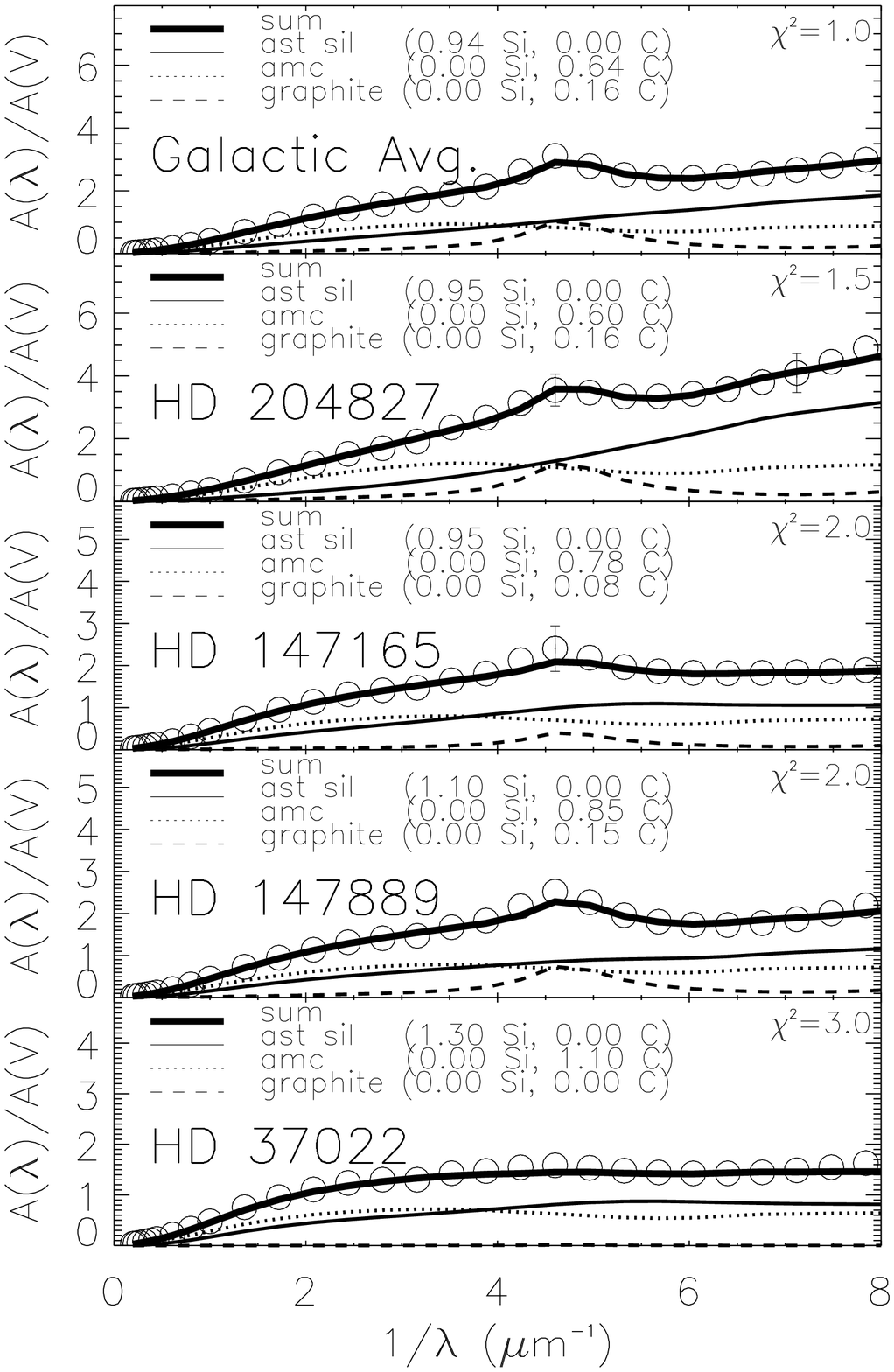}{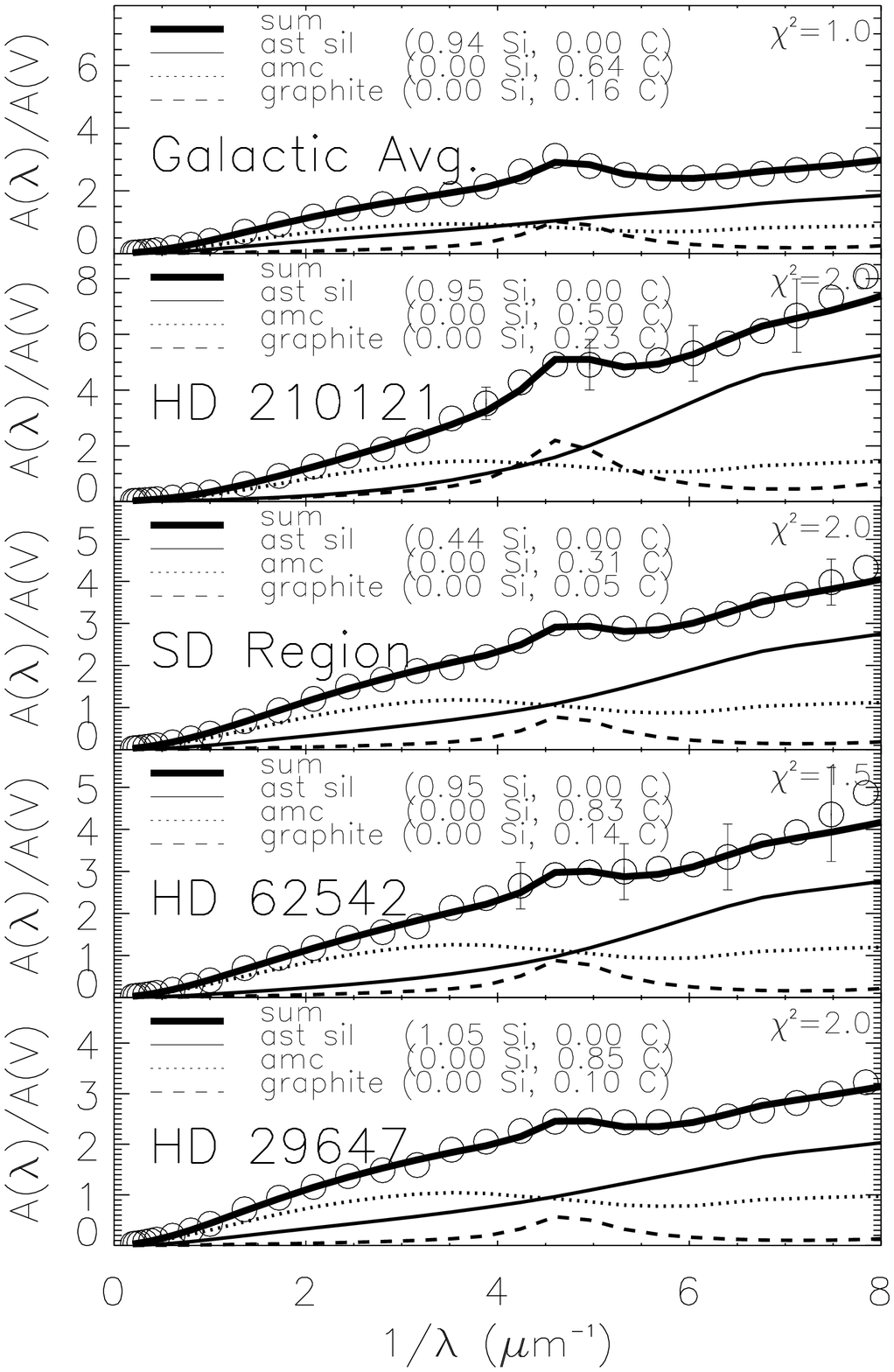}
\vspace{0.5 in}
\caption{Three-component extinction models for Galactic sightlines.
Each panel shows the model fit to the extinction curve, including
the contribution of each component.  The fraction of the ``cosmic'' Si
and C (amorphous carbon and graphite) 
utilized is listed in the figure legend (See Table 1 for abundances).} 
\label{fig-1}
\end{figure*}

\begin{figure*}
\epsscale{1.0}
\plottwo{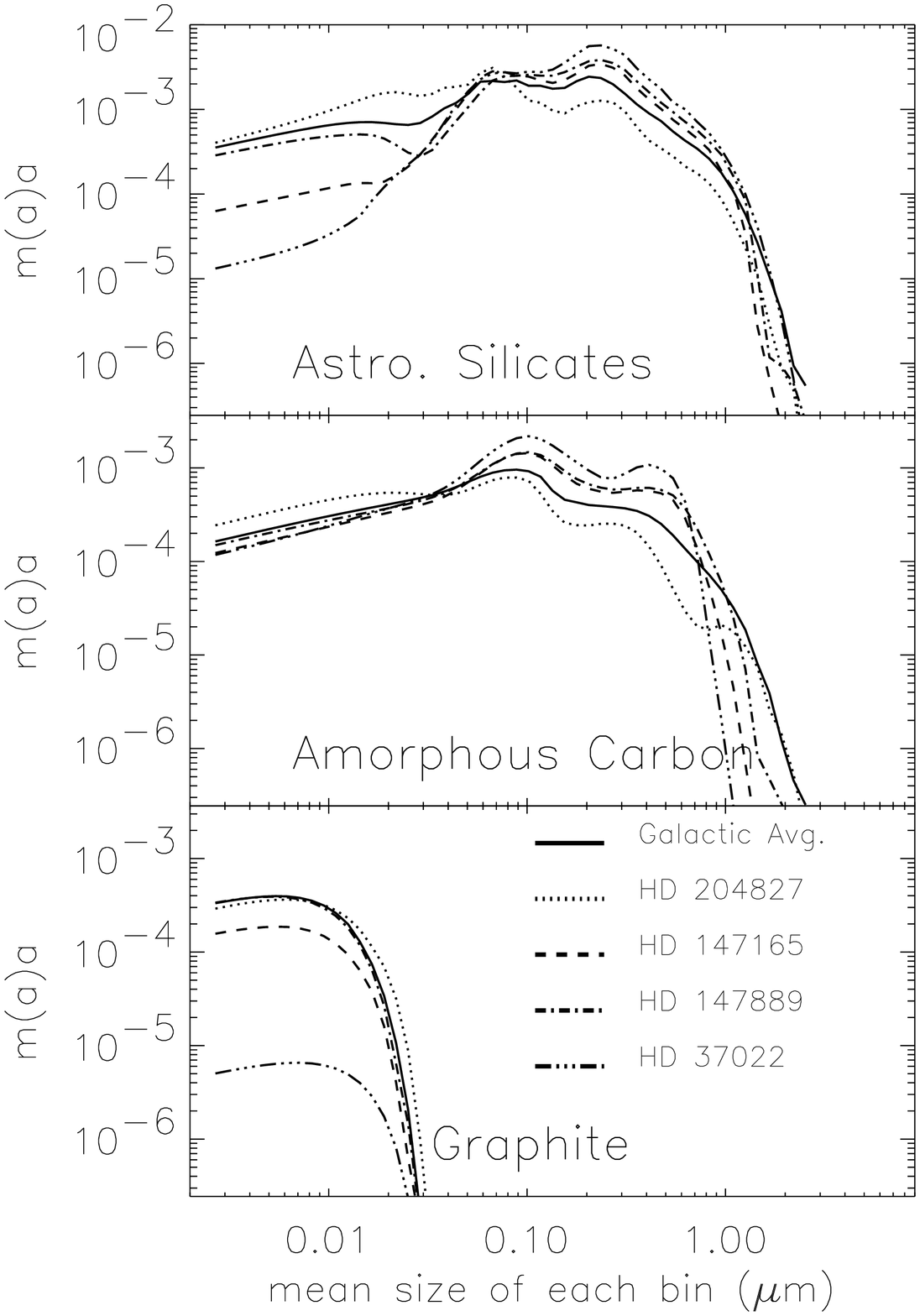}{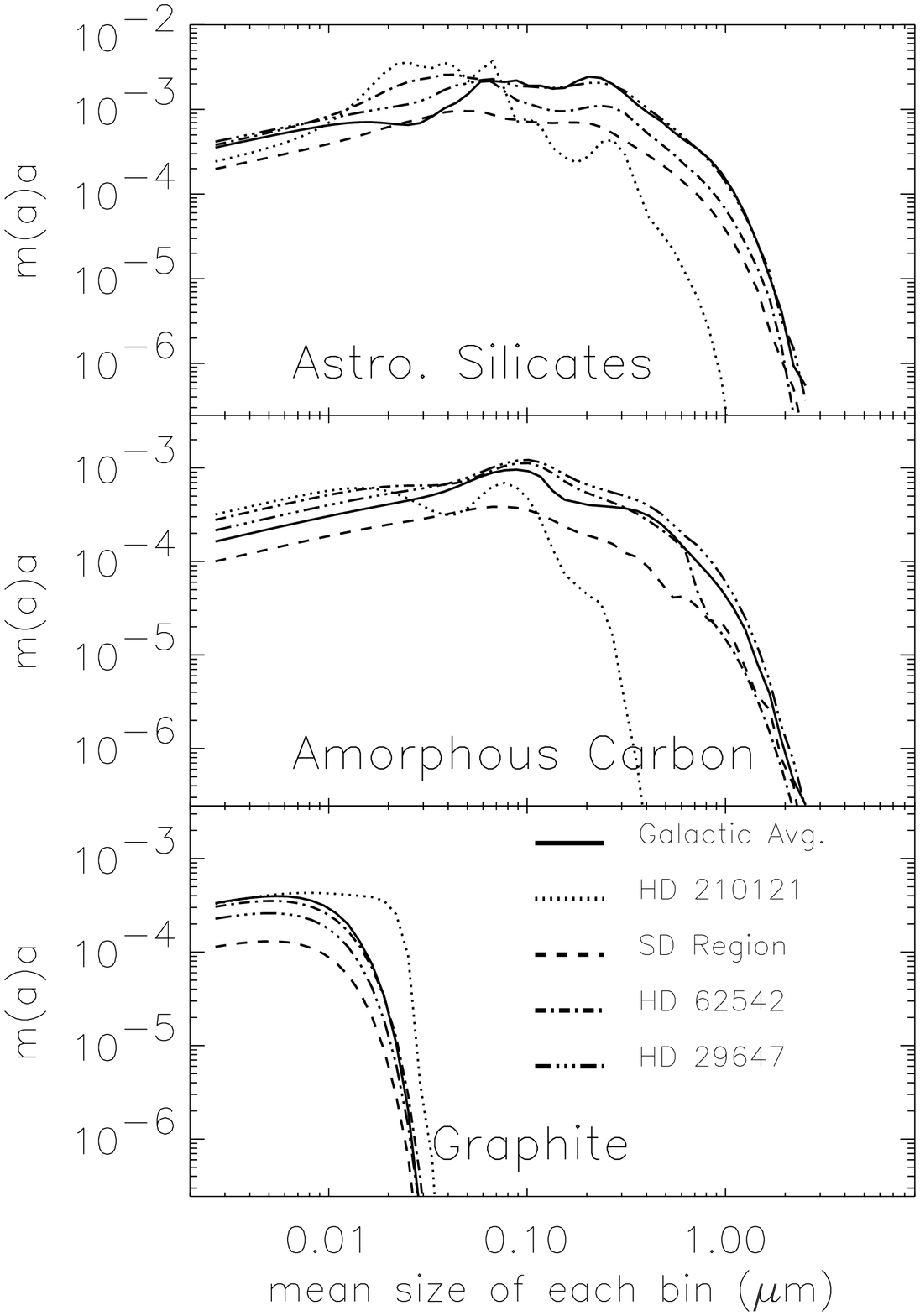}
\vspace{0.5 in}
\caption{Three-component extinction models for Galactic sightlines.
Each panel contains the resultant mass distributions relative to the
mass of hydrogen.  For the three-component model, amorphous carbon 
replaces graphite as primary source of carbonaceous continuum extinction. 
In order to minimize the graphite requirement in the three-component model,
we adopt a new template function; size bins
are included only through 0.04 \micron.} \label{fig-2}
\end{figure*}

\begin{figure*}
\epsscale{1.0}
\plottwo{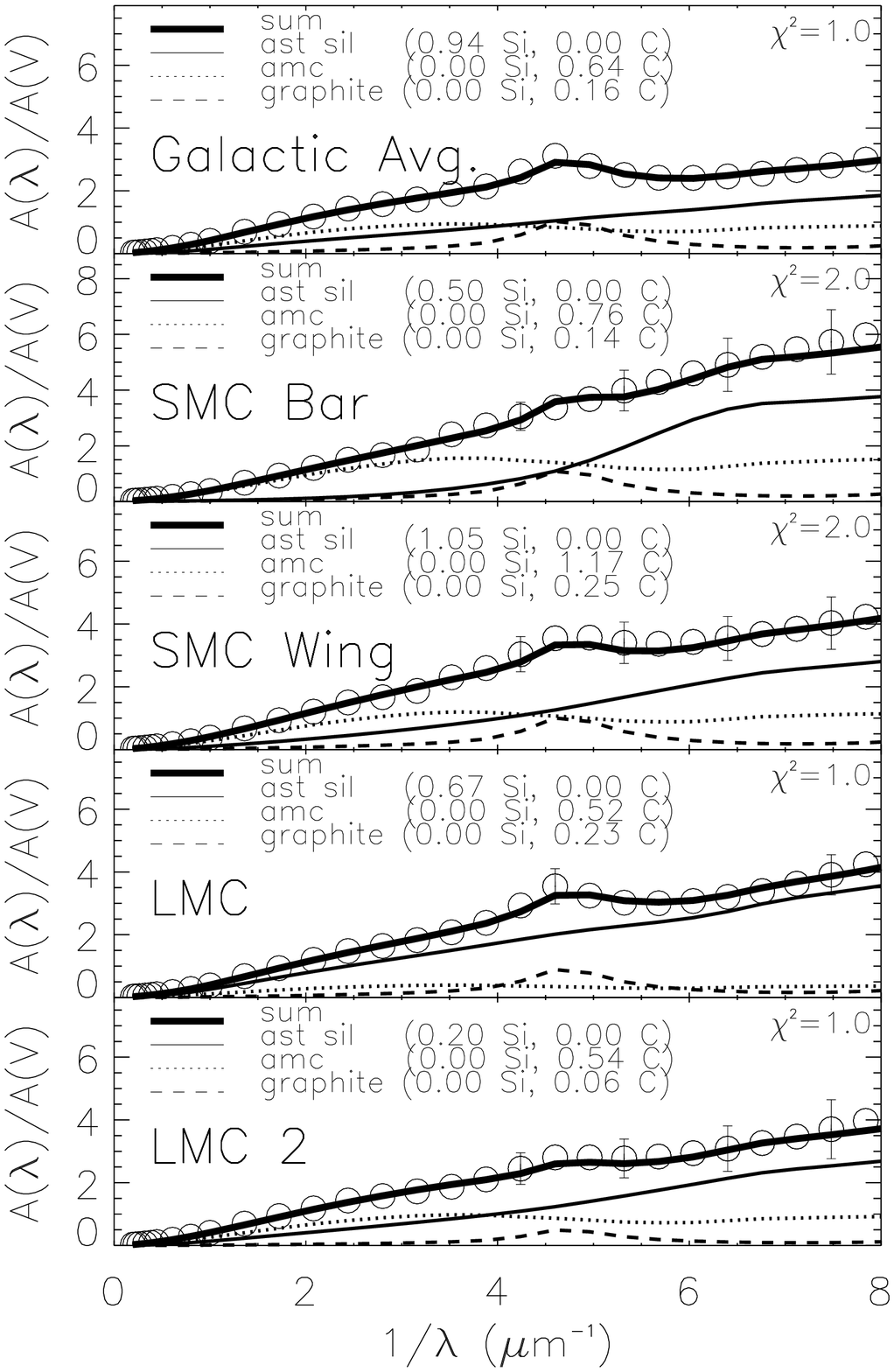}{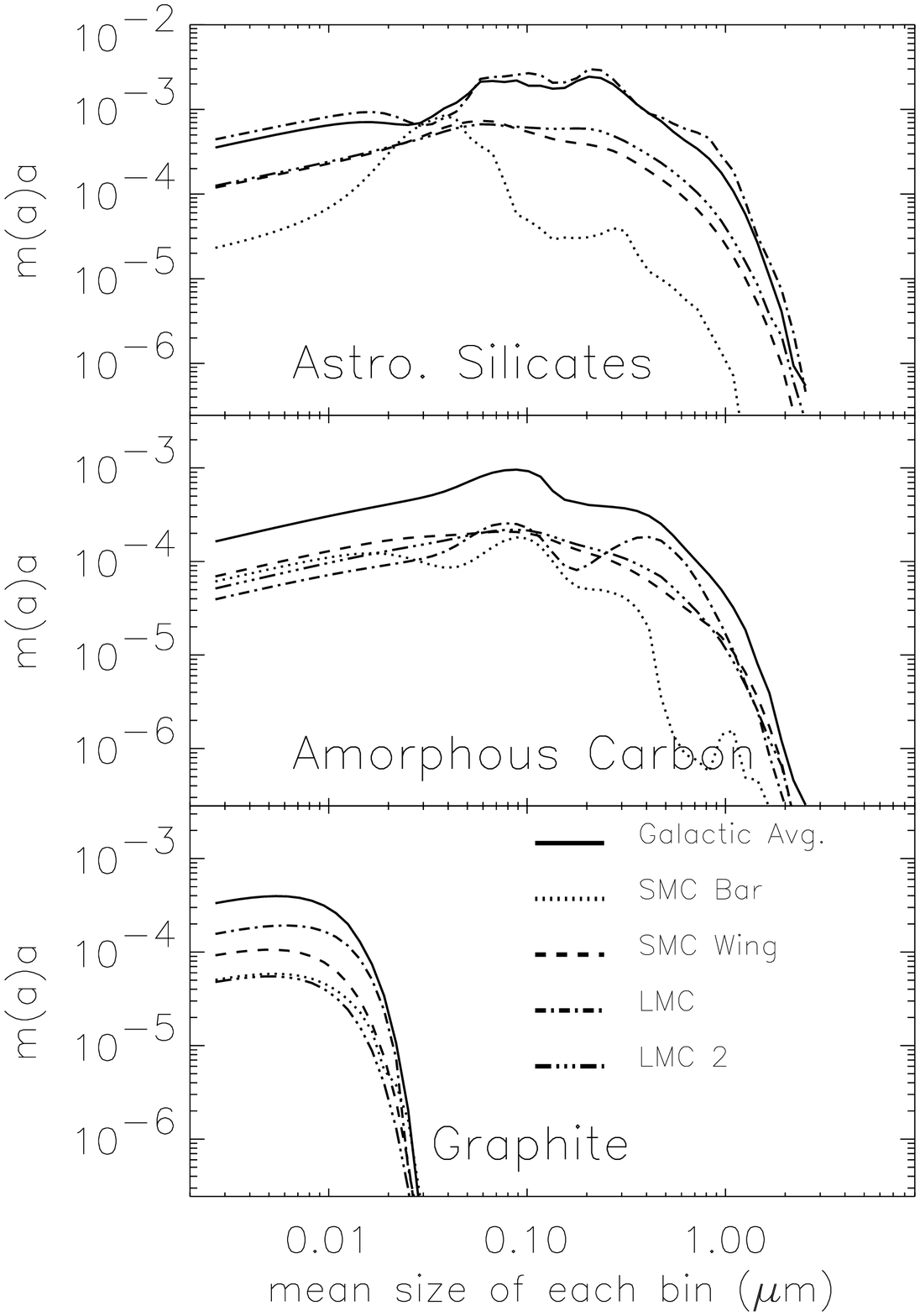}
\vspace{0.5 in}
\caption{Three-component extinction modes for Magellanic Cloud sightlines.
The left-hand panel shows the model fits to the extinction curves, including
the contribution of each component.
The right-hand panel contains the resultant mass distributions relative to the
mass of hydrogen. Plotted in the same way as Figures 1 and 2.} \label{fig-3}
\end{figure*}

\begin{figure*}
\epsscale{1.0}
\plotone{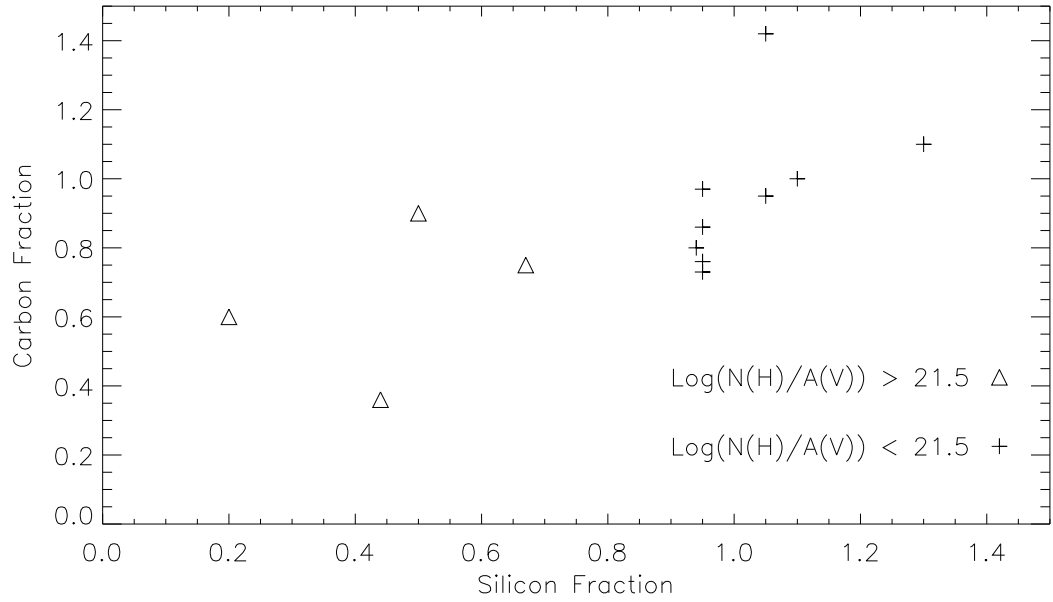}
\vspace{0.5 in}
\caption{Fraction of available silicon used plotted against fraction of available
carbon (amorphous carbon + 
graphite) used for the sightlines in our sample. The sample has been
divided into those with large and small values of the gas-to-dust ratio.} 
\label{fig-4}
\end{figure*}

\begin{figure*}
\epsscale{1.0}
\plottwo{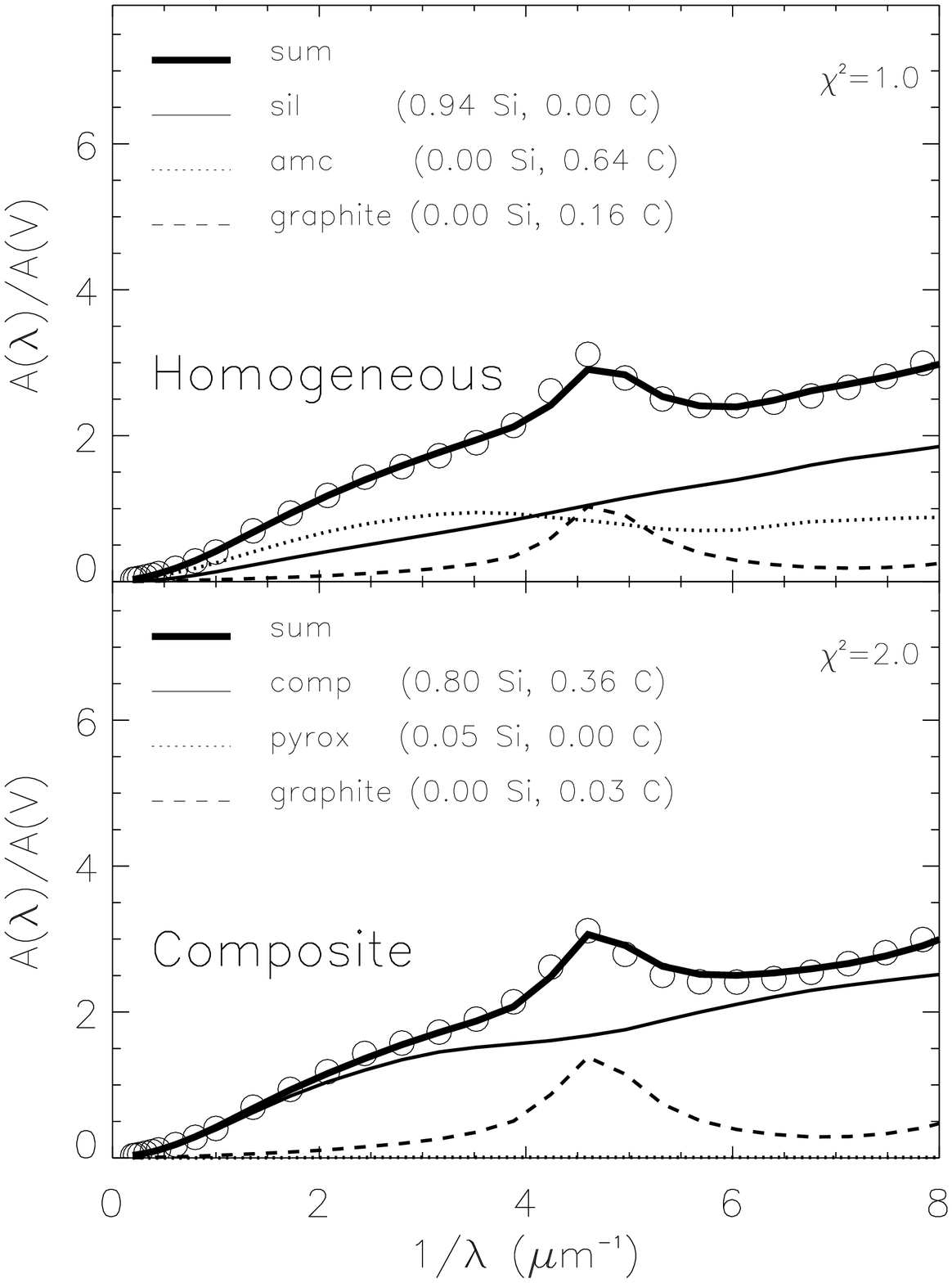}{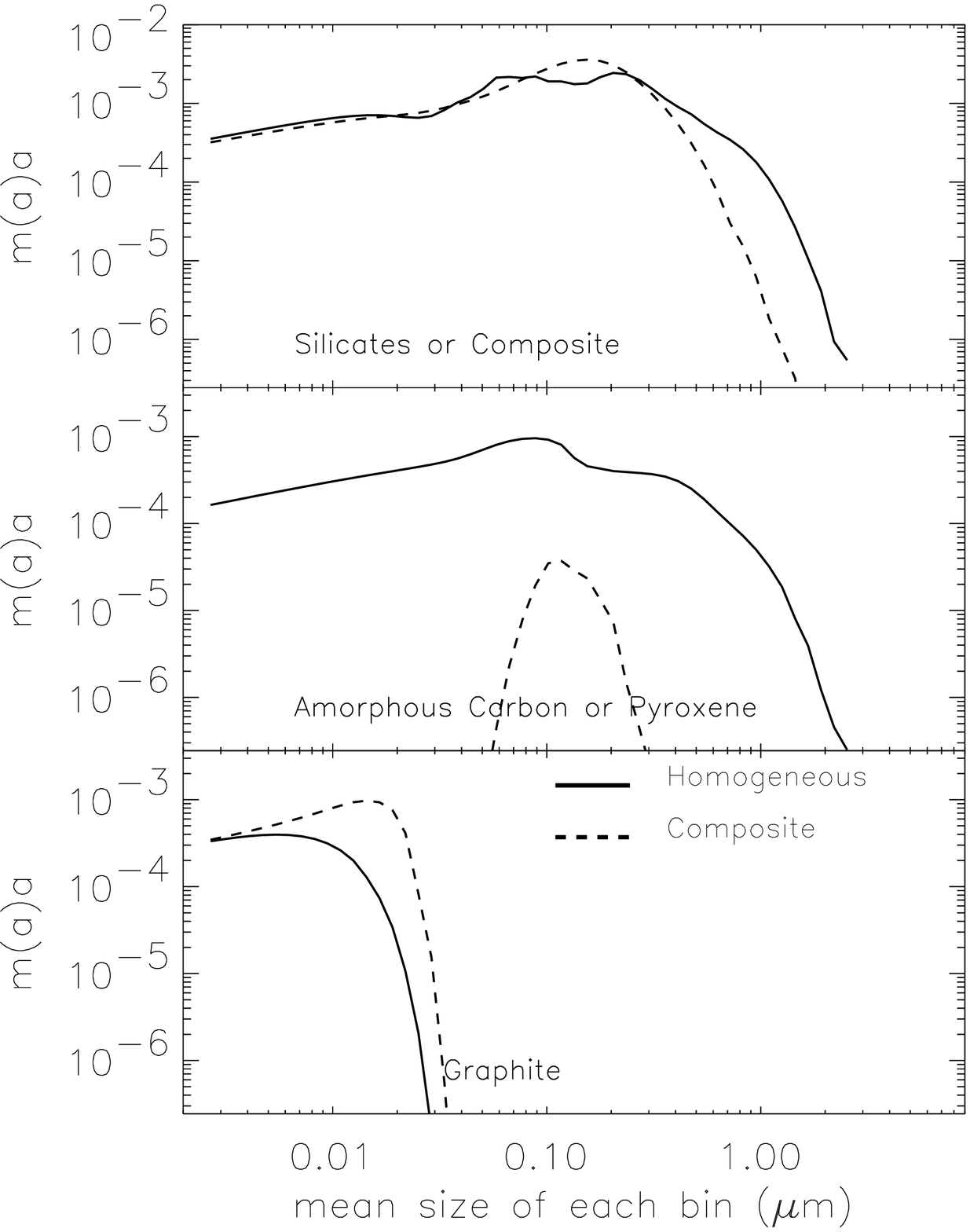}
\vspace{0.5 in}
\caption{Composite model including pyroxene and oxides 
for the average Galactic extinction compared to the 
3-component
fit used in Figures 1 and 2.  The left-hand panel shows the model fits to 
the extinction curves, including
the contribution of each component.
The right-hand panel contains the resultant mass distributions relative to the
mass of hydrogen. Plotted in the same way as Figures 1 and 2.} \label{fig-5}
\end{figure*}


\begin{references}
\reference{} Aannestad, P.A. 1992, ApJ, 443, 653
\reference{} Biermann, P. \& Harwit, M. 1980, ApJ, 241, L105
\reference{} Bohlin, R. C., Savage, B. D., \& Drake, J. F. 1978, \apj, 224, 132
\reference{} Bouchet, P., Lequeux, J., Maurice, E., Prevot, L., \& Prevot-
Burnichon, M.L. 1985,
A\&A, 149, 330
\reference{} Cardelli, J.A., \& Clayton, G.C. 1991, AJ, 101, 1021
\reference{} Cardelli, J. A., Clayton, G. C., \& Mathis, J. S. 1989, \apj,
345, 245
\reference{} Cardelli J.A., Mathis, J.S., Ebbets, D.C., \& Savage, B.D. 1993, 
ApJ, 402, L17
\reference{} Cardelli, J. A., \& Savage, B.D. 1988, ApJ, 325, 864
\reference{} Cardelli, J.A., Sembach, K.R., \& Mathis, J.S. 1992, AJ, 104, 1916
\reference{} Cardelli J.A., Sofia, U.J., Savage, B.D., Keenan, F.P., \& 
Dufton, 
P.L. 1994, ApJ, 420, L29
\reference{} Clayton, G.C., Gordon, K.D., and Wolff, M.J. 2000, ApJS, 129, 147
\reference{} Clayton, G.C., \& Hanson. M.M. 1993, A.J., 105,1880
\reference{} Clayton, G. C., Wolff, M. J., Allen, R. G., \& Lupie, O. L.
1995, \apj, 445, 947
\reference{} Clayton, G. C., Wolff, M. J., Gordon, K.D., \& Misselt, K.A. 2000,
ASP Conf. Ser., 196, 41
\reference{} Desert, F.-X., Boulanger, F., \& Puget, J.L. 1990, A\&A, 237, 215
\reference{} Dorschner J., Begemann B., Henning Th., Jaeger, C., \& 
Mutschke, H. 1995, A\&A, 300, 503
\reference{} Draine, B.T., \& Anderson, N. 1985, ApJ, 292, 494
\reference{} Draine, B.T., \& Lee, H.M. 1984, ApJ, 285, 89
\reference{} Dwek, E. et al. 1997, ApJ, 475, 565
\reference{} Fitzpatrick, E. L. 1985, \apj, 299, 219
\reference{} Fitzpatrick, E. L. \& Massa, D. 1990, \apjsupp, 72, 163
\reference{} Gillett, F.C., \& Forrest, W.J. 1973, ApJ, 179, 483 
\reference{} Gordon, K. D. \& Clayton, G.C. 1998, \apj, 500, 816
\reference{} Gordon, K. D., Clayton, G.C., Misselt, K.A., Wolff, M.J., \& 
Landolt, A.U. 2002, ApJ, Submitted
\reference{} Greenberg, J.M. 1968, in Nebulae and Interstellar Matter, ed. 
B.M. Middlehurst and L.H. Aller (Chicago: University of Chicago Press), p. 221
\reference{} Greenberg, J.M., \& Chlewicki, G. 1983, ApJ, 272, 563
\reference{} Grevesse,N., \& Anders, E. 1989, in Cosmic Abundances of Matter, AIP 
Conf. Proc., 183, 1
\reference{} Gunderson, K. A., Clayton, G. C., \& Green, J. 1998,
\pasp, 110, 60
\reference{} Hayakawa, H., \& Hayakawa, S. 1988, PASJ, 40, 341
\reference{} Henning Th., Begemann B., Mutschke H., \& Dorschner J. ,1995,
A\&AS, 112, 143
\reference{} Jenniskens, P. 1993, A\&A, 274, 653
\reference{} Kim, S.-H. \& Martin, P. G. 1995, \apj, 442, 172
\reference{} Kim, S.-H. \& Martin, P. G. 1996, \apj, 462, 296
\reference{} Kim, S.-H., Martin, P. G., \& Hendry, P.D. 1994, \apj, 422, 164
\reference{} Koornneef, J. 1982, A\&A, 107, 247
\reference{} Koornneef, J. 1983, A\&A, 128, 84
\reference{} Laor, A. \& Draine, B. T. 1993, \apj, 402, 441
\reference{} Larson, K.A., Whittet, D.C.B., \& Hough, J.H. 1996, ApJ, 472, 755
\reference{} Larson, K.A., Wolff, M.J., Roberge, W.G., Whittet, D.C.B., \& 
He, L. 2000, ApJ, 532, 1021
\reference{} Li, A., \& Draine, B.T. 2001, ApJ, 554, 778
\reference{} Li, A., \& Greenberg, J.M. 1997, A\&A, 323, 566
\reference{} Li, A., \& Greenberg, J.M. 1997, A\&A,339, 591
\reference{} Martin. N., Maurice, E., \& Lequeux, J. 1985, A\&A, 215, 219
\reference{} Massa, D., \& Fitzpatrick, E. L. 2000, ApJS, 126, 517
\reference{} Massa, D., Savage, B. D., \& Fitzpatrick, E. L. 1983, \apj, 266, 662
\reference{} Mathis, J. S. 1996, ApJ, 472, 643
\reference{} Mathis, J. S. 1998, \apj, 497, 824
\reference{} Mathis, J. S. 2000, JGR, 105, 10269
\reference{} Mathis, J. S., \& Cardelli 1992, ApJ, 398, 610
\reference{} Mathis, J. S., Rumpl, W. \& Nordsieck, K.H. 1977, ApJ, 217, 425
\reference{} Mathis, J. S., \& Wallenhorst, S.G. 1981, ApJ, 244, 483
\reference{} Mathis, J. S., \& Whiffen, G. 1989, ApJ, 341, 808
\reference{} Misselt, K. A., Clayton, G. C., \& Gordon, K. A. 1999, \apj,
512, 128
\reference{} Morgan, D.H., \& Nandy, K. 1982, MNRAS, 199, 979
\reference{} Nandy, K.,  Morgan, D.H.,  \& Houziaux, L. 1984 , MNRAS, 211, 895
\reference{} O'Donnell, J.E., \& Mathis, J.S 1997, ApJ, 479, 806
\reference{} Oort, J.H., \& van de Hulst, H.C. 1946, BAN, 10, 187
\reference{} Ossenkopf, V. 1991, A\&A, 251, 210
\reference{} Pei, Y. C. 1992, \apj, 395, 130
\reference{} Rieke, G.H., \& Lebofsky, M.J. 1985, ApJ, 288, 618
\reference{} Rouleau \& Martin 1991, \apj, 377, 526
\reference{} Russell, S. C. \& Dopita, M. A. 1992, \apj, 384, 508
\reference{} Savage, B.D., Drake, J.F., Buditch, W., \& Bohlin, R.C. 1977, 
ApJ, 
215, 788
\reference{} Savage, B.D., \& Sembach, K.R. 1996, ARA\&A, 34, 279
\reference{} Sembach, K.R., \& Danks, A.C. 1994, A\&A, 289, 539
\reference{} Snook, K.~J., 1999, Ph.D Thesis, Stanford University, 
Stanford, California
\reference{} Snow, T.P., \& Witt, A.N. 1996, ApJ, 468, L65\\
\reference{} Sofia, U.J., Cardelli, J.A. \& Savage, B.D. 1994, ApJ, 430, 650\\
\reference{} Sofia, U.J., \& Meyer, D.M. 2001, ApJ, 554, L221
\reference{} Spitzer, L.,  \& Fitzpatrick, E.L. 1993, ApJ, 409, 299\\
\reference{} Stecher, T.P., \& Donn, B. 1965, ApJ, 142, 1681
\reference{} Valencic, L. et al. 2002, in preparation
\reference{} Venn, K.A. 1999, ApJ, 518, 405
\reference{} Warren, S., 1984, Appl. Opt., 23, 1206
\reference{} Weingartner, J.C., \& Draine, B.T. 2001, ApJ, 548, 296
\reference{} Welty, D. E., Lauroesch, J.T., Blades, J.C., Hobbs, L.M., \& York, 
D.G. 1997, ApJ, 489, 672
\reference{} Welty, D. E., Lauroesch, J.T., Blades, J.C., Hobbs, L.M., \& York, 
D.G. 2001, ApJ, 554, L75
\reference{} Whittet, D.C.B. 1992, Dust in the Galactic Environment, (IOP: 
Bristol)
\reference{} Whittet, D.C.B., Gerakines, P.A., Hough, J.H., \& Shenoy, S.S. 2001, 
ApJ, 547, 872
\reference{} Whittet, D.C.B., Martin, P.G., Fitzpatrick, E.L., \& Massa, D. 1993, 
ApJ, 408, 573
\reference{} Woolf, N.J., \& Ney, E.P. 1969, ApJ, 155, L181
\reference{} Zubko, V.G. 1999, ApJ, 513, L29
\reference{} Zubko, V.G., Krelowski, J., \& Wegner, W. 1996a, MNRAS, 283, 577
\reference{} Zubko, V.G., Krelowski, J., \& Wegner, W. 1998, MNRAS, 294, 548
\reference{} Zubko, V.G., Mennella, V., Colangeli, L., \&
Bussoletti, E.  1996b, MNRAS, 282, 1321
\end{references}
\end{document}